\documentclass[%
reprint,
 amsmath,amssymb,
 aps,
]{revtex4-1}

\usepackage{graphicx}
\usepackage{graphics}
\usepackage{adjustbox} 
\usepackage{dcolumn}
\usepackage{bm}
\usepackage{dsfont}
\usepackage{amsthm}
\usepackage[dvipsnames]{xcolor}
\usepackage{comment}
\usepackage{apptools}
\AtAppendix{\counterwithin{lemma}{section}}
\usepackage{mathtools}
\usepackage{floatrow} 
\DeclareFloatFont{tiny}{\tiny}
\usepackage[T1]{fontenc}

\theoremstyle{definition}

\theoremstyle{definition}

\def\ket#1{\left\vert #1 \right\rangle}
\def\bra#1{\left\langle #1 \right\vert}

\newcommand{\be}{\begin{equation}}
\newcommand{\ee}{\end{equation}}
\newcommand{\bp}{\begin{pmatrix}}
\newcommand{\ep}{\end{pmatrix}}
\newcommand{\ben}{\begin{enumerate}}
\newcommand{\een}{\end{enumerate}}

\begin{document}

\title{Benchmarking near-term quantum devices with the Variational Quantum Eigensolver and the Lipkin-Meshkov-Glick model}

\author{Kenneth Robbins}
\affiliation{Department of Physics and Astronomy, Tufts University, Medford MA
}%
\author{Peter J. Love}
 \altaffiliation[Also at ]{Brookhaven National Laboratory}
\affiliation{Department of Physics and Astronomy, Tufts University, Medford MA
}%

\begin{abstract}
 The Variational Quantum Eigensolver (VQE) is a promising algorithm for Noisy Intermediate Scale Quantum (NISQ) computation. Verification and validation of NISQ algorithms' performance on NISQ devices is an important task. We consider the exactly-diagonalizable Lipkin-Meshkov-Glick (LMG) model as a candidate for benchmarking NISQ computers. We use the Bethe ansatz to construct eigenstates of the trigonometric LMG model using quantum circuits inspired by the LMG's underlying algebraic structure. We construct circuits with depth $\mathcal{O}(N)$ and $\mathcal{O}(\log_2N)$ that can prepare any trigonometric LMG eigenstate of $N$ particles. The number of gates required for both circuits is $\mathcal{O}(N)$. The energies of the eigenstates can then be measured and compared to the exactly-known answers.
\end{abstract}

\pacs{Valid PACS appear here} 
\maketitle

\section{Introduction}\label{intro}

Quantum simulation is a leading application of quantum computation. Quantum Phase Estimation (QPE) can find Hamiltonian eigenvalues but is challenging to implement in the Noisy Intermediate Scale Quantum (NISQ) era of computation due to a high resource cost~\cite{Preskill2018quantumcomputingin,PhysRevX.6.031007,bharti2021noisy}. The Variational Quantum Eigensolver (VQE), a quantum-classical hybrid algorithm which requires fewer resources, is more suited to current NISQ computers which as yet lack fault-tolerance and fully-coherent evolution~\cite{Peruzzo_2014,cerezo2020variational}.

Consider an ansatz state $\ket{\psi\left(\Theta_M\right)}$ and a Hamiltonian $H=\sum_k H_k$ where $\Theta_M\equiv\{\theta_1...\theta_M\}$ is a set of ansatz parameters. VQE uses a quantum processing unit to estimate the expectation value $\bra{\psi\left(\Theta_M\right)}H_k\ket{\psi\left(\Theta_M\right)}$ for all $k$ and outputs this information to a classical machine. Said classical machine then sums these expectation values to estimate $\bra{\psi\left(\Theta_M\right)}H\ket{\psi\left(\Theta_M\right)}$ and alters the ansatz parameters $\Theta_M$ according to some optimization algorithm to minimize the estimated value. This process is iterated until a minimum is obtained: a variational estimate of the ground-state energy~\cite{Peruzzo_2014}.

Problems with implementing VQE include overcoming noisy devices, barren plateaus in the phase space and the choice of a suitable ansatz~\cite{cerezo2020variational}. This paper concerns itself, however, with providing a way to benchmark an individual quantum machine's performance using exactly-solvable models and VQE. This is done by providing both an exactly-solvable model and a suitable ansatz-preparation circuit for all relevant cases of the model. Further, we discuss ways to obtain the ansatz parameters $\Theta_M$ which will prepare given LMG eigenstates and give examples of this.

In the case of an exactly-solvable Hamiltonian we can find the eigenvalues and eigenstates analytically and therefore these models are useful as benchmarks for VQE because we can compare VQE outputs to exact calculations. This requires the construction of a circuit that prepares the exact eigenstates. In the present paper we investigate the Lipkin-Meshkov-Glick (LMG) model which describes two nuclear shells~\cite{LIPKIN1965188,pan1999analytical,ORTIZ2005421,LermaH.:2013cla,cervia2020exactly}. The Bethe ansatz suggests a circuit structure valid for any number of qubits; constructing these circuits is the main topic of the present paper.

Previous work discussed quantum simulation of exactly-solvable Hamiltonians such as the LMG for applications such as benchmarking or as a testbed for exotic matter detection~\cite{cerezo2020variational,dallairedemers2020application,pleff,plgates}. In particular, the Bethe ansatz has been considered to design state preparation circuits of polynomial depth for correlated fermionic states and the $XXZ$ spin chain~\cite{vandyke2021preparing,pldepth}. Exact models have also been theoretically explored for non-variational quantum simulation~\cite{Cervera_Lierta_2018,Verstraete_2009,Schmoll_2017}.

In Section~\ref{lmgsec} we introduce the LMG model and explore its underlying Gaudin algebra structure. In Section~\ref{secego} we apply that structure to construct circuits that prepare LMG eigenstates. In Section~\ref{exsec} we provide specific examples of how to connect the exact eigenstates to our proposed circuits for low numbers of particles. We discuss our findings in Section~\ref{discsec}.

\begin{table*}\label{usefultab} 
 \begin{adjustbox}{max width=\textwidth}
 \begin{tabular}{||c | c | c ||}
 \hline
 Quantity & Domain & Description   \\ [0.5ex] 
 \hline\hline
 $N=2M+\nu_a+\nu_b$ & $\mathbb{Z}$ & Number of simulated particles \\
 \hline
 $M$ & $\mathbb{Z}$ & Number of ansatz Parameters $=$ Particle Pairs  \\ 
 \hline
$\nu_a,\nu_b$ & $\{0,1\}\subset\mathbb{Z}$ & Fiducial state occupation numbers \\
\hline
$V,W$ & $\mathbb{R}$ & Interaction Strengths\\
\hline
 $g\left(V,W,N\right),\eta\left(V,W\right)$ & $\mathbb{R}$ & Model Parameters\\
 \hline
 $E_\ell$ & $\mathbb{C}$ & Spectral Parameters (Pairing/Pairon Energies) \\
 \hline
 $S^\pm_m, S^z_m$ & GGA & $XXZ$ Gaudin Algebra Generators \\
 \hline
$K^\pm_\lambda, K^z_\lambda$ & $su(1,1)\oplus su(1,1)$ & Describe the Gaudin Algebra \\
 \hline
 $\lambda,\tau$ & $\{1,2\}\subset\mathbb{Z}$ & Indices over the Modes \\
 \hline
 $m,\ell,n$ & $\{1...M\}\subset\mathbb{Z}$ & Indices over the Pairs \\
 \hline
  $\ket{M,\nu_a,\nu_b;j}$ & Fock Space & Bosonic Fock State ($1\leq j\leq M+1$) \\
 \hline
   $\ket{\psi_{M}}$ & Circuit Space & Circuit Output State after $M$ Iterations \\
 \hline
$c_j$ & $\mathbb{C}$ & Generic vector coefficients such that $\sum |c_j|^2=1$\\
\hline
\end{tabular}
\end{adjustbox}
\caption{A table of relevant variables, operators and indices used throughout the paper. When setting up an Eigenstate Generating Operator (EGO), one must specify $N$ and solve the Bethe equations with all allowed values of $M,\nu_a,\nu_b$ in order to be able to generate all eigenstates for the given $N$. For the solutions to the Bethe Equations~(\ref{beqs}) $E_\ell$ it is known that $E_\ell\in \mathbb{R}$ in the trigonometric LMG model $(V^2 > W^2)$ and for at least one case in the hyperbolic LMG model $(V^2 < W^2)$~\cite{LermaH.:2013cla}. This paper deals exclusively with the cases when the spectral parameters $E_\ell$ are real for all $1\leq \ell\leq M$.}
\end{table*}

\section{The Lipkin-Meshkov-Glick (LMG) Model}\label{lmgsec}
This section details the representations and algebraic qualities of the LMG model so that we may use them later to construct an LMG eigenstate-preparation circuit. To that end, we focus on the bosonic representation of the model and its relationship to generalized Gaudin-Richardson models.

\subsection{The Hamiltonian and Transformations}

The LMG model describes a collection of $N$ fermions and was introduced by S. Fallieros to study the nuclear phenomenon of giant monopole resonance \cite{LIPKIN1965188,FallierosThesis}. In 1965 Lipkin, Meshkov and Glick proposed using the model to benchmark ``techniques and formalisms for treating many-body systems.'' In the LMG model each fermion can be in one of two $N$-fold degenerate energy levels labeled by quantum number $\sigma=\pm 1$; the original physical example given is two total-angular-momentum coupling shells of a nucleus. Another quantum number $1\leq k\leq N$ labels fermionic modes~\cite{LIPKIN1965188}. The LMG model is an example of a pairing model. Pairing models such as the LMG and the Bardeen-Cooper-Schrieffer (BCS) have long been relevant in nuclear and condensed matter physics, and as such they are of interest as targets for quantum simulation~\cite{wu2002polynomial}.

The LMG Hamiltonian is given by

\begin{multline}\label{fham}
    H=\frac{1}{2}\sum_{k,\sigma}\sigma f^\dag_{k\sigma}f_{k\sigma}+\frac{V}{2N}\sum_{k,k',\sigma}f^\dag_{k\sigma}f^\dag_{k'\sigma}f_{k'-\sigma}f_{k-\sigma}\\
    +\frac{W}{2N}\sum_{k,k',\sigma}f^\dag_{k\sigma}f^\dag_{k'-\sigma}f_{k'\sigma}f_{k-\sigma}\\
\end{multline}
where fermionic operator $f_{k\sigma}^\dag$ creates a fermion in the $k\sigma$ state, $f_{k\sigma}$ annihilates a particle in the state defined by mode label $k$ and energy level $\sigma$. $V,W\in\mathbb{R}$ are interaction strengths. The energy gap between the levels is set to unity without loss of generality by rescaling $V$ and $W$. The LMG model can be rewritten in terms of angular momentum operators~\cite{ORTIZ2005421}:
    \begin{equation}
        J_\pm\equiv\sum_{k=1}^N f^\dag_{k\pm}f_{k\mp}\text{ ,}\quad J_z\equiv\frac{1}{2}\sum_{k=1}^N\sum_{\sigma=\pm}\sigma f^\dag_{k\sigma}c_{k\sigma}.
    \end{equation}
The total angular momentum quantum number is fixed as $j=N/2$. In terms of these operators, the LMG Hamiltonian is:

\begin{equation}\label{angham}
 H=J_z+\frac{V}{2N}( J_+ J_++ J_- J_-)+\frac{W}{2N}( J_+ J_-+ J_- J_+).
\end{equation}

The angular momentum operators can be written in terms of bosonic operators via the Jordan-Schwinger mapping~\cite{ORTIZ2005421, LermaH.:2013cla, osti_4389568}.
\begin{equation}
    J_+\equiv b^\dag a\quad J_-\equiv a^\dag b\quad J_z\equiv\frac{1}{2}\left(n_b-n_a\right)
\end{equation}
where operators $a^{(\dag)}, b^{(\dag)}$ annihilate (create) bosons in modes $a$ or $b$, respectively and $n_b\equiv b^\dag b$ and  $n_a\equiv a^\dag a$ are their number operators. Mode $a$ corresponds to particles in the lower energy level in the fermionic representation, mode $b$ to those in the higher and the total number of quanta $N_{\text{fermion}}=N_{\text{boson}}\equiv N$. We will use the following representation of the LMG Hamiltonian:

\begin{multline}\label{bham}
H=\frac{n_b-n_a}{2}+\frac{V}{2N}\left(b^\dag b^\dag aa+ a^\dag a^\dag bb\right)\\
+\frac{W}{N}\left(\frac{n_a+n_b}{2}+n_an_b\right).
\end{multline}
Note that if $V=0$ the system is diagonal in the bosonic Fock basis.

\subsection{Algebraic Structure}

The LMG model is a $2$-level $XXZ$ Gaudin model~\cite{ORTIZ2005421,LermaH.:2013cla,pan1999analytical}. This means that the Hamiltonian~(\ref{bham}) can be written in terms of the constants of motion of an $XXZ$ Gaudin algebra~\cite{ORTIZ2005421, LermaH.:2013cla, pan1999analytical}. We now define the terms used in the following form for the Gaudin Hamiltonian~\cite{ORTIZ2005421}:
\begin{equation}
\begin{multlined}\label{genham}
    H=\sum_\lambda \epsilon_\lambda K^z_\lambda +\sum_{\lambda,\tau,\lambda\neq\tau}\left(\epsilon_\lambda-\epsilon_\tau\right)\times{}\\
    \Big(X_{\lambda\tau}\left(K^+_\lambda K^-_\tau+K^-_\lambda K^+_\tau\right)
    -2Z_{\lambda\tau}K^z_\lambda K^z_\tau\Big).\\
\end{multlined}
\end{equation}
Indices $\lambda,\tau$ run from $1$ to $2$ and $\epsilon_2-\epsilon_1$ is the energy gap between the two levels given in the fermionic description of the LMG Hamiltonian, which we set to unity without loss of generality. Operators $K^\pm_\lambda$ and $K^z_\lambda$ are generators of $su(1,1)\oplus su(1,1)$ and can be expressed in terms of bosonic operators:
\begin{subequations}\label{kops}
    \begin{equation}
        K^+_1=\frac{1}{2}a^\dag a^\dag \quad\quad K^+_2=\frac{1}{2}b^\dag b^\dag,
    \end{equation}
    \begin{equation}
        K^-_1=\frac{1}{2}aa\quad\quad K^-_2=\frac{1}{2}bb,
    \end{equation}
    \begin{equation}
        K^z_1=\frac{1}{2}\left(n_a+\frac{1}{2}\right)\quad\quad K^z_2=\frac{1}{2}\left(n_b+\frac{1}{2}\right)
    \end{equation}
\end{subequations}
obeying the commutators
\begin{equation}\label{11com}
    \left[K_\lambda^+,K_\tau^-\right]=-2\delta_{\lambda\tau}K^z_\tau \quad\quad \left[K^z_\lambda,K^\pm_\tau\right]=\pm K^\pm_\tau.
\end{equation}
With these we can construct the Gaudin operators $S^\pm_m$ and $S^z_m$~\cite{ORTIZ2005421,LermaH.:2013cla}:
\begin{subequations}
\begin{equation}
    S^+_m=X_{m1}K^+_1+X_{m2}K^+_2=\left(X_{m1}a^\dag a^\dag+X_{m2}b^\dag b^\dag\right)/2
\end{equation}
\begin{equation}
    S^-_m=-X_{m1}K^-_1-X_{m2}K^-_2=-\left(X_{m1}aa+X_{m2}bb\right)/2
\end{equation}
\begin{multline}
    S^z_m=-\frac{1}{2}-Z_{m1}K^z_1-Z_{m2}K^z_2\\
    =-\frac{1}{2}\left(1+Z_{m1}\left(a^\dag a+\frac{1}{2}\right)+Z_{m2}\left(b^\dag b+\frac{1}{2}\right)\right)\\
\end{multline}
\end{subequations}
where $X_{m\ell}=-X_{\ell m}$ and $Z_{m\ell}=-Z_{\ell m}$ are matrices with indices $m,\ell$ running from $1$ to integer $M$, the number of particle pairs. It is these antisymmetric matrices that give the $XXZ$ Gaudin algebra its name. These operators obey the commutation relations that define the $XXZ$ Gaudin algebra:
\begin{subequations}\label{GGA2}
\begin{equation}\label{GGA2a}
    \left[S^\pm_m,S^\pm_\ell\right]=0
\end{equation}
\begin{equation}\label{GGA2b}
    \left[S^-_m,S^+_\ell\right]=\begin{cases}
    -2X_{m\ell}\left(S^z_m-S^z_\ell\right) & m\neq \ell \\
    2f_m\frac{\partial S^z_m}{\partial E_m} & m=\ell
    \end{cases}
\end{equation}
\begin{equation}\label{GGA2c}
    \left[S^z_m,S^\pm_\ell\right]=\begin{cases}
    \pm\left(X_{m\ell}S^\pm_m-Z_{m\ell}S^\pm_\ell\right) & m\neq \ell \\
    \mp f_m\frac{\partial S^\pm_m}{\partial E_m} & m=\ell
    \end{cases}
\end{equation}
\end{subequations}
where $f_m$ can be written as one holomorphic function divided by another~\cite{ORTIZ2005421}. We use Equations~(\ref{kops}-\ref{GGA2}) to rewrite Equation~(\ref{genham}) as
\begin{multline}
    H=\frac{\epsilon_2-\epsilon_1}{2}\left(n_b-n_a\right)+\frac{X_{12}}{2}\left(a^\dag a^\dag bb+b^\dag b^\dag aa\right)\\
    -Z_{12}\left(n_an_b+\frac{n_b+n_a}{2}\right)\\
\end{multline}
up to a constant additive term of $(1-Z_{12})/4$. Comparing this to the LMG Hamiltonian given in Equation~(\ref{bham}) we see that $X_{12}=V/N$ and $Z_{12}=-W/N$. This Gaudin framework enables us construct the eigenstates in Section~\ref{secego}.

\section{Eigenstate Generating Operator}\label{secego}
In the context of the LMG Hamiltonian, the Bethe ansatz outputs two relevant equations: one that can characterize all LMG eigenstates in terms of a single operator and another that characterizes all of the eigenvalues. We define the former as an Eigenstate Generating Operator (EGO).

\subsection{Fock Space Operator}
The LMG Hamiltonian~(\ref{bham}) is exactly diagonalizable via the Bethe ansatz~\cite{LIPKIN1965188,ORTIZ2005421,pan1999analytical}. We follow Reference~\cite{ORTIZ2005421} and define the EGO as

\begin{equation}\label{ego}
    \prod^M_{\ell=1}\left(\frac{(a^\dag)^2}{E_\ell+\eta}+\frac{(b^\dag)^2}{E_\ell-\eta}\right).
\end{equation}

Application of the EGO to a fiducial state $\ket{\nu_a,\nu_b}$ results in an (unnormalized) LMG eigenstate. Note that the EGO~(\ref{ego}) has $M+1$ terms after applying the product because $a^\dag$ and $b^\dag$ commute. As all eigenstates of the LMG can be written in terms of the EGO, any LMG eigenstate $\ket{M,\nu_a,\nu_b;j}$ can be written as
\begin{equation}\label{linfockspace}
    \ket{M,\nu_a,\nu_b;j}=\sum_{k=0}^M c_{k} \ket{N-\nu_b-2k,2k+\nu_b}
\end{equation}
where $c_{k}$ are normalized coefficients unique to the $j$\textsuperscript{th} eigenstate that can be efficiently classically generated with a given $M,\nu_a,\nu_b$. The size of the support for LMG is therefore linear in $M$, which suggests that its eigenstates can be generated with generalized $\ket{W}$ preparation circuits~\cite{Cruz_2019}. 

The variables $\nu_a,\nu_b$ are restricted to be $0$ or $1$ and represent unpaired particles while $M$ is the number of particle pairs. This restriction on $\nu_a,\nu_b$ allows for four fiducial states: $\ket{0,0}, \ket{0,1}, \ket{1,0}$ and $\ket{1,1}$. The integer $M$ is also the requisite number of iterations of the factor $G\left(E_\ell\right)$, and is set by~\cite{LermaH.:2013cla,ORTIZ2005421}
\begin{equation}\label{ndef}
    N=2M+\nu_a+\nu_b.
\end{equation}
The variables $E_\ell$ are spectral parameters, also known as pair energies or pairons~\cite{LermaH.:2013cla}. The spectral parameters are obtained by solving the Bethe ansatz equation(s)~\cite{ORTIZ2005421,LermaH.:2013cla}:
\begin{multline}\label{beqs}
    1-\frac{\eta}{N\left(E_\ell^2-\eta^2\right)}\bigg[g N\left(\nu_a-\nu_b\right)\left(1+sE_\ell^2\right)\\
    +2VE_\ell\left(1+\nu_a+\nu_b\right)\bigg]+2g\sum^M_{n\neq l=1}\frac{1+sE_\ell E_n}{E_\ell-E_n}=0.
\end{multline}
The variables $g$ and $\eta$ are analytic functions of interaction strengths $V$ and $W$ parameter~\cite{pan1999analytical,ORTIZ2005421}:
\begin{subequations}
\begin{equation}
    g=\sqrt{\frac{V^2-W^2}{sN^2}},
\end{equation}
\begin{equation}
    \eta=-\sqrt{\frac{V+W}{s\left(V-W\right)}},
\end{equation}
\end{subequations}
where $s$ is a sign parameter
\begin{equation}
    s\equiv
    \begin{cases}
     +1 & \text{if $V^2>W^2$}\\
     0 & \text{if $V^2=W^2$}\\ 
     -1 & \text{if $V^2<W^2$}\\
    \end{cases}.
\end{equation}

As a guide for circuit design, the EGO also illustrates the $\mathcal{O}(N/2)$ vector support and suggests that a linear-gate-count circuit should exist. Both of these qualities make the LMG model suitable for benchmarking in NISQ devices because it can then be simulated with low resources in both gate- and qubit-counts.

For a given $M,\nu_a,\nu_b$ there are $M+1$ distinct solutions for the spectral parameters. Substituting these into the EGO then results in $M+1$ unnormalized eigenstates which are themselves superpositions of up to $M+1$ Fock basis states. To generate all the eigenstates for a given particle number $N$, one must solve the Bethe equations for all possible combinations of $M,\nu_a,\nu_b$ and apply all distinct solution sets for the spectral parameters to the EGO, of which there should be a total of $N+1$.

Given an output of the EGO as a function of the spectral parameters with $M$ applications of the EGO factor, it is possible to find the normalized $(M+1)$\textsuperscript{th}-level output vector in terms of the spectral parameters as well. We do so for the case of $\nu_a=\nu_b=\nu$ and $W=0\neq V$. Let $\ket{\text{EGO}_{M,\nu}}$ be the (unnormalized) $M$-level output vector from fiducial state $\ket{\nu,\nu}$:
\begin{equation}
    \ket{\text{EGO}_{M,\nu}}\equiv \sum_{j=0}^M d_j\ket{2M+\nu-2j,\nu+2j}.
\end{equation}
For cases where $\nu_a\neq \nu_b$ we will use $\ket{\text{EGO}_{M,\nu_a,\nu_b}}$. After applying the EGO factor to $\ket{\text{EGO}_{M,\nu}}$ one more time and normalizing we have:
\begin{multline}\label{m1veccalc}
    \ket{M+1,\nu,\nu;k}=\gamma^{-1/2}\sum_{j=0}^Md_j\times\\
    \Bigg(\frac{\prod_{i=1}^2\sqrt{w_{ji}}\ket{w_{j2},x_{j0}}}{E_{M+1}-1}\\
    +\frac{\prod_{i=1}^2\sqrt{x_{ji}}\ket{w_{j0},x_{j2}}}{E_{M+1}+1}\Bigg)
\end{multline}
where
\begin{equation}
    w_{ji}\equiv 2M+\nu-2j+i\quad\quad x_{ji}\equiv \nu+2j+i
\end{equation}
and $\gamma$ is a normalization factor
\begin{multline}
    \gamma\equiv\sum_{m=0}^Md_m^2\frac{\prod_{i=1}^2w_{mi}}{\left(E_{M+1}-1\right)^2}+\frac{\prod_{i=1}^2x_{mi}}{\left(E_{M+1}+1\right)^2}\\
    +2\sum_{n=0}^{M-1}d_nd_{n+1}\frac{\prod_{i=1}^2\sqrt{w_{n,i-2}x_{ni}}}{E_{M+1}^2-1}.
\end{multline}

Therefore given an eigenstate $\ket{M,\nu,\nu;j}$ or an unnormalized $\ket{\text{EGO}_{M,\nu}}$, both of which are functions of $\{E_1...E_M\}$, it is possible to calculate the form of $\ket{M+1,\nu,\nu,k}$ as a function of $\{E_1...E_{M+1}\}$.

The trigonometric regime of the LMG defined by $V^2> W^2$ guarantees real spectral parameters~\cite{LermaH.:2013cla} which, in turn, guarantees eigenstates with real coefficients. Conversely, in the hyperbolic regime defined by $W^2> V^2$ for some values of $N,V,W$ complex coefficients can occur. The circuits proposed within this paper assume the trigonometric case but are also valid in the hyperbolic if $E_\ell\in\mathbb{R}$ for all $1\leq\ell\leq M$.

The Bethe ansatz also allows us to compute the eigenvalues $\omega$ in terms of the spectral parameters $E_\ell$~\cite{ORTIZ2005421,LermaH.:2013cla}:
\begin{multline}\label{evaleq} 
    \omega=\frac{W\left(\nu_a+\nu_b+2\nu_a\nu_b\right)+N\left(\nu_b-\nu_a\right)}{2N}\\
    -\frac{\eta}{N}\sum_{\ell=1}^M\frac{ gN\left(1+\nu_a+\nu_b\right)\left(1+s E_\ell^2\right)-2V\left(\nu_b-\nu_a\right)E_\ell}{E_\ell^2-\eta^2}.\\
\end{multline}



Solving the Bethe Equations~(\ref{beqs}) simplifies in special cases. If one sets $W=0\neq V$ and $N$ is even (where $\nu_a=\nu_b\equiv \nu$), then $g\rightarrow V/N, \eta\rightarrow -1, s\rightarrow 1$. Thus the Bethe equations simplify in that case:
\begin{equation}\label{simpbeqs}
    1+\frac{2\left(1+2\nu\right)VE_\ell}{N\left(E_\ell^2-1\right)}
    +2\frac{V}{N}\sum_{n\neq \ell=1}^M\frac{1+E_\ell E_n}{E_\ell-E_n}=0.
\end{equation}

For $M=1$ and $2$ we can find analytical solutions for the Bethe equations. For $M=1$ the simplified Bethe equations become
\begin{multline}\label{simpm1}
    1+V\frac{1+2\nu}{\left(1+\nu\right)\left(E_1^2-1\right)}E_1=0\\
    =E_1^2+V\frac{1+2\nu}{1+\nu}E_1-1.\\
\end{multline}
This is a quadratic equation with respect to $E_1$. The same is found for the full $M=1$ Bethe equation(s) but with differing coefficients. We may solve Equation~(\ref{simpm1}) to yield
\begin{equation}
    E_1=-\frac{V\left(1+2\nu\right)\pm\sqrt{V^2\left(1+2\nu\right)^2+N^2}}{N}.
\end{equation}

The simplified $M=2$ Bethe equations are
\begin{subequations}
\begin{equation}
    1+V\frac{\left(1+2\nu\right)E_1}{\left(2+\nu\right)\left(E_1^2-1\right)}+V\frac{1+E_1E_2}{\left(2+\nu\right)\left(E_1-E_2\right)}=0,
\end{equation}
\begin{equation}
 1+V\frac{\left(1+2\nu\right)E_2}{\left(2+\nu\right)\left(E_2^2-1\right)}+V\frac{1+E_1E_2}{\left(2+\nu\right)\left(E_2-E_1\right)}=0.
\end{equation}
\end{subequations}
Multiplying through by the denominators gives quartic polynomials; the same is found for the full $M=2$ Bethe equations. Two sets of solutions are
\begin{equation}\label{m2ell}
    E_\ell=-\frac{V(1+2\nu)\pm(-1)^\ell\sqrt{4(1+\nu)^2+V^2(1+2\nu)^2}}{2(1+\nu)}.
\end{equation}

Beyond $M=2$, the addition of more terms with $E_\ell-E_n$ in their denominators leads to an increase in polynomial degree of the Bethe equations. For a given nonzero $M>1$, the degree of the Bethe equations is $M+2$. Therefore the Bethe equations for $M\geq 3$ are at least quintic in degree and lack analytical radical solutions in the general case~\cite{abel_2012}. As such we assume for the rest of the paper that all solutions for the Bethe equations beyond $M=2$ are obtained numerically. If necessary, the LMG Hamiltonian~(\ref{angham}) can be written as an $(N+1)\times(N+1)$ matrix and diagonalized, which gives the problem a complexity of $\mathcal{O}(N^3)$.

\subsection{Eigenstate Preparation Circuits}\label{wnsec}
The $\ket{W_M}$ state is defined as 
\begin{equation}
    \ket{W_M}\equiv\frac{1}{\sqrt{M}}\left(\sum_{j=1}^M X_j\right)\ket{0}^{\otimes M}=\sum_{j=0}^{M-1}\frac{\ket{2^j}_{M}}{\sqrt{M}}
\end{equation}
where $X_j$ is the Pauli-$X$ gate acting on qubit $j$~\cite{vidal3qubs}. We also introduce the notation $\ket{n}_m$ for a computational state representing integer $n$ as a bitstring of length $m$. The $\ket{W_M}$ states have support linear in $M$ and uniform real amplitudes. Previous work has found $\ket{W_M}$ preparation circuits which have gate counts $\mathcal{O}(M)$ and depth $\mathcal{O}(\log_2M)$~\cite{Cruz_2019}. These circuits can be adapted into EGO circuits for the LMG Hamiltonian~(\ref{bham}).

It has been shown that the LMG ground and quantum critical states can be prepared with high fidelity using $\mathcal{O}(1)$ iterations of the Variational Quantum-Classical Simulation (VQCS) protocol~\cite{Ho_2019}. This method does not use the exactly-solvable nature of the LMG model via the Bethe ansatz. Our circuit uses the Fock-space-to-qubit encoding
\begin{equation}\label{sampleencoding}
    \ket{2^{k}}_{M+1}\iff \ket{2M+\nu_a-2k,\nu_b+2k}
\end{equation}
where $k$ is an integer from $0$ to $M$. Each EGO factor is given by a controlled-$R_y$ gate and a controlled-NOT (CNOT) gate. The EGO circuit is given by
\begin{equation}\label{logego} 
    \prod_{\ell=1}^M\text{C}^{M+2-\ell}\left(\text{NOT}_{f\left(M+1-\ell\right)}\right)\text{C}^{f\left(M+1-\ell\right)}\left(R_y\left(\theta\right)_{M+2-\ell}\right)
\end{equation}
where $\text{C}^m\left(Q_n\right)$ designates a controlled-$Q$ gate with the control on qubit $m$ and gate $Q$ is acting on qubit $n$. The function $f(n)$ controls the depth of the circuit:
\begin{equation}\label{fnseq}
    f(n)\equiv\begin{cases} n \quad \text{for depth }\mathcal{O}(2M)\\
    n-2^{\lfloor \log_2 n\rfloor}+1\quad\text{for depth }\mathcal{O}(\log_22M)\\ 
    \end{cases}
\end{equation}
where $n$ is a positive integer.

\begin{figure*}
    \centering
    \includegraphics[scale=.4]{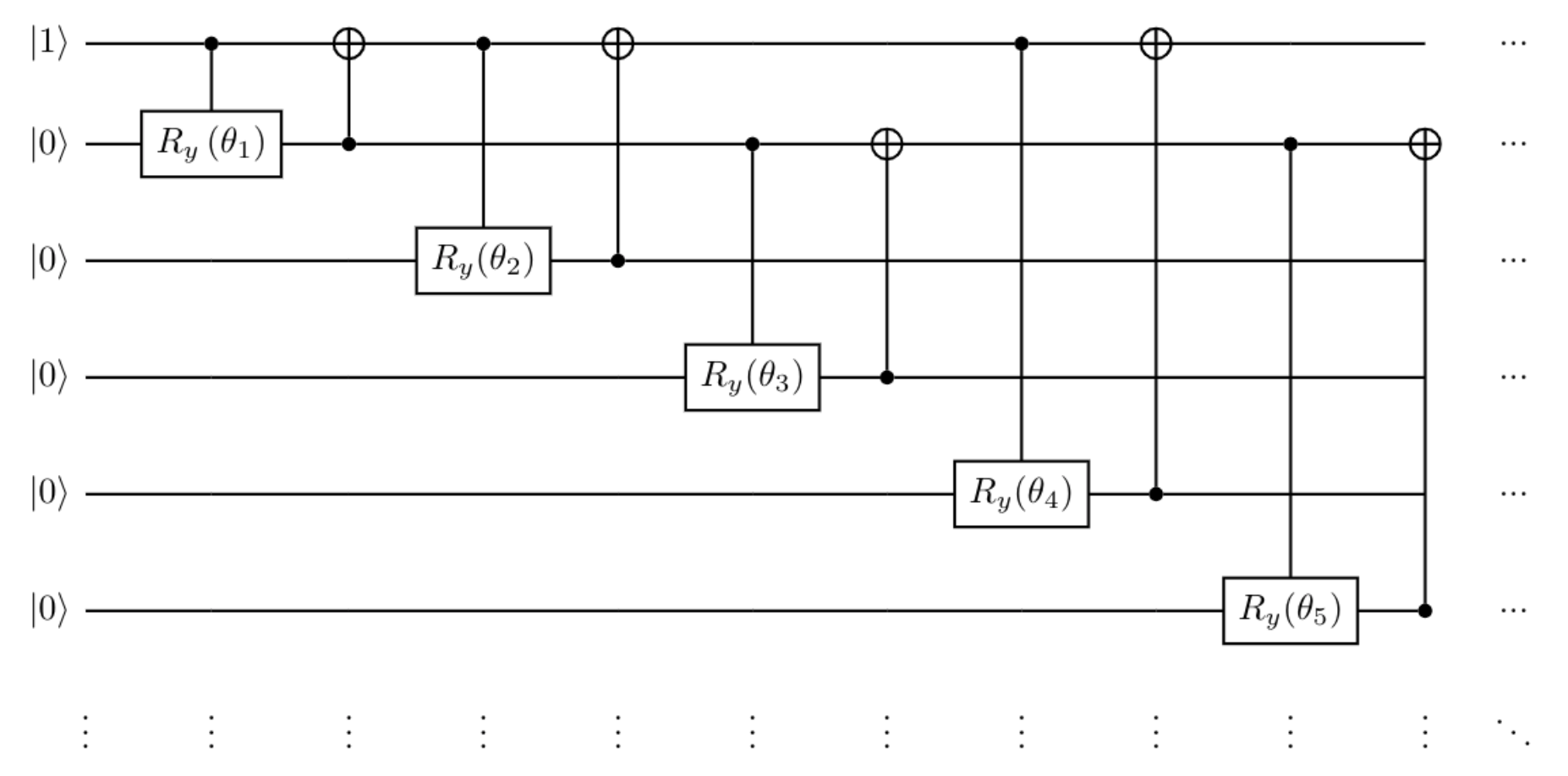}
    \caption{An eigenstate generating operator (EGO) circuit for the $N$-particle Lipkin-Meshkov-Glick (LMG) model. Its form is given by Equations~(\ref{logego}) and~(\ref{fnseq}). An alternative form with linear circuit depth is also explored within the paper. Each controlled-$R_y$ and CNOT gate pair takes the place of one factor in the EGO in Equation~(\ref{ego}). By varying the ansatz parameters $\Theta_M\equiv \{\theta_1,...\theta_M\}$ (where $M\approx N/2$ is given by Equation~(\ref{ndef})) any eigenstate of the trigonometric ($V^2 > W^2$) LMG can be set up on a quantum machine using $M+1$ qubits. Some cases of $V^2\leq W^2$ LMG eigenstates can also be prepared. This circuit requires $2M$ two-qubit gates but has time complexity $\mathcal{O}(\log_2 M)$. This is allowed because the first $2^0$ factor commutes with itself, the next $2^1$ factors commute with each other, as do the next $2^2$ and so on. This circuit is adapted from previously-established $\ket{W_M}$-state preparation circuits~\cite{Cruz_2019}. The fiducial input state is always $\ket{1}\otimes\ket{0}^{\otimes M}$.}
    \label{linfig}
\end{figure*}

The set of all rotation angles is denoted $\Theta_M$. The EGO circuit in Equation~(\ref{logego}), also shown in Figure~\ref{linfig}, acts on fiducial state $\ket{1}\otimes\ket{0}^{\otimes M}$ on $M+1$ qubits. The logarithmic time complexity comes from parallelizing blocks of gates; the first $2^0=1$ gate should be run alone, then the next $2^1$ can be run in parallel, as can the next $2^2$ and so on. 

For the linear-depth circuit, the output vector is the following:
\begin{multline}
    \ket{\psi_M}=\prod_{j=1}^M\sin\frac{\theta_j}{2}\ket{2^0}_{M+1}\\
    +\sum_{\ell=1}^M\cos\frac{\theta_{\ell}}{2}\prod_{k=1}^{\ell-1}\sin\frac{\theta_k}{2}\ket{2^{M+1-\ell}}_{M+1}.
\end{multline}
This parameterization is identical to $(M+1)$-dimensional spherical coordinates up to a swap in the first two components~\cite{nspherestuff}. Thus the output state $\ket{\psi_M}$ of the linear-depth EGO circuit parameterizes the $M$-sphere over the given support, so it is a functional ansatz for any trigonometric LMG eigenstate.

The linear-depth circuit allows for simpler angle specification than the logarithmic-depth circuit. For a generic real unit $(M+1)$-vector $\ket{T_M}$ over the specified circuit's support
\begin{equation}
    \ket{T_M}=\sum_{j=1}^{M+1}c_j\ket{2^{j-1}}_{M+1}
\end{equation}
the linear-depth circuit will output target vector $\ket{T_M}$ where the angles are given by~\cite{nspherestuff}:

\begin{equation}\label{linearangles}
    \theta_j=
    \begin{cases}
    2\cos^{-1}q_{M,j} & \text{if }j\leq M\\
    2\text{ sgn}\left(c_1\right)(\cos^{-1}q_{M,j}-\pi)+2\pi & \text{if }j=M
    \end{cases}
\end{equation}
where we define $q_{M,j}$ as
\begin{equation}\label{quotient}
    q_{M,j}\equiv \frac{c_{M+2-j}}{\sqrt{\sum_{k=1}^{M+2-j}c_k^2}}.
\end{equation}

The coefficients $c_j$ can be determined in terms of $E_\ell$ from Equation~(\ref{m1veccalc}). If $E_\ell$ are known they can be substituted directly into the EGO~(\ref{ego}), normalized and the resultant LMG eigenstate can be directly prepared with angles given by Equations~(\ref{linearangles}). 

If $c_{M+2-j}\ll \sqrt{\sum_{k=1}^{M+2-j}c^2_k}$ then the inverse cosine in Equation~(\ref{linearangles}) will be approximately equal to $\pi/2$. In that case for $j<M$ then $\theta_j\approx \pi$ while $\theta_M\approx 0$. This behavior is explored in Appendix~\ref{excircs} for the case of $N=20, M=10$. 

We prove the logarithmic circuit's generality for $M\geq 2$ in Appendix~\ref{logproofsec} by demonstrating that the circuit can reproduce any real combination of coefficients for a unit state vector of dimension $M+1$ and Hamming weight $1$. Example output states for both the logarithmic-depth and linear-depth circuits are given in Appendix~\ref{logputs}.

\section{Examples}\label{exsec}
The EGO, given by Equation~(\ref{ego}), is an operator on two bosonic modes. The quantum circuit described in Figure~\ref{linfig} and Equation~(\ref{logego}) is an operator on a $2^{M+1}$ dimensional Hilbert space. To aid in relating the two we provide examples for small numbers of particles. We use the logarithmic-depth circuit for these demonstrations, though in principle the linear-depth would also function. The angles for the linear-depth circuit are also given in Equation~(\ref{linearangles}).

\subsection{$\mathbf{N=1}$ Eigenstates}\label{n1sec}

For the case of $N=1$ we are forced by Equation~(\ref{ndef}) to consider only $M=0$ while either $\nu_a$ or $\nu_b$ must be $1$ but not both. As $M=0$ the EGO is the identity operator; therefore $\ket{0,1}$ and $\ket{1,0}$ are the $N=1$ eigenstates. The EGO circuit for this case is the identity. Eigenstates will be labeled as $\ket{M,\nu_a,\nu_b;j}$ where index $j$ can vary from $1$ to $M+1$ and labels the individual eigenstates for a given $N$:
\begin{equation}
    \ket{0,0,1;1}=\ket{0,1}\quad\quad\ket{0,1,0;2}=\ket{1,0}.
\end{equation}

\subsection{$\mathbf{N=2}$ Eigenstates}
The case of $N=2$ in Equation~(\ref{ndef}) allows $M=0$ with $\nu_a=\nu_b=1$ and $M=1$ with $\nu_a=\nu_b=0$. We have already seen that the $M=0$ EGO is the identity in Section~\ref{n1sec} so we take $\ket{1,1}$ as the eigenstate $\ket{0,1,1;1}$. For $M=1$ and any $N$ we have the unnormalized eigenstate from the EGO as
\begin{multline}\label{e1eq}
    \ket{\text{EGO}_{1,\nu_a,\nu_b}}=\frac{\sqrt{2-\left(-1\right)^{\nu_a}}}{E_1+\eta}\ket{\nu_a+2,\nu_b}\\
    +\frac{\sqrt{2-\left(-1\right)^{\nu_b}}}{E_1-\eta}\ket{\nu_a,\nu_b+2}
\end{multline}
with an omitted overall factor of $\sqrt{2}$. To normalize $\ket{\text{EGO}_{1,\nu_a,\nu_b}}$ multiply by a normalization factor
\begin{equation}
    \gamma=\frac{\left(E_1^2-\eta^2\right)}{\sqrt{4\left(E_1^2+\eta^2\right)-\left(-1\right)^{\nu_a}\left(E_1-\eta\right)^2-\left(-1\right)^{\nu_b}\left(E_1+\eta\right)^2}}.
\end{equation}

The $M=1,N=2$ Bethe equation is
\begin{equation}\label{beqs2}
    1-\frac{V\eta E_1}{E_1^2-\eta^2}=E_1^2-V\eta E_1-\eta^2=0
\end{equation}
which is quadratic in $E_1$. Therefore the solution is
\begin{equation}\label{n2spec}
    E_{1}=\frac{\eta}{2}\left(V\pm\sqrt{V^2+4}\right).
\end{equation}
For $M=1$ the Bethe equations are relatively simple. Substitution of Equation~(\ref{n2spec}) into Equation~(\ref{e1eq}) yields two unnormalized eigenstates $\ket{1,0,0;1}$ and $\ket{1,0,0;2}$. The $N=2$ eigenstates are:
\begin{subequations}\label{n2kets}
\begin{equation}
    \ket{1,0,0;1}=\gamma_1\left(\ket{2,0}-\frac{2+\sqrt{4+V^2}}{V}\ket{0,2}\right)
\end{equation}
\begin{equation}
    \ket{1,0,0;2}=\gamma_2\left(-\ket{2,0}+\frac{2-\sqrt{4+V^2}}{V}\ket{0,2}\right)
\end{equation}
\end{subequations}
where the normalization factor $\gamma_j$ is equal to
\begin{equation}\label{normn2}
    \gamma_j=\left(1+\left(\frac{-2+\left(-1\right)^j\sqrt{4+V^2}}{V}\right)^2\right)^{-1/2}.
\end{equation}

For both the linear and logarithmic $M=1$ EGO circuit, the output state is
\begin{multline}\label{out1}
    \ket{\psi_{1,0}}=\sin\frac{\theta_1}{2}\ket{01}+\cos\frac{\theta_1}{2}\ket{10}\\
    \equiv \sin\frac{\theta_1}{2}\ket{2^0}_2+\cos\frac{\theta_1}{2}\ket{2^1}_2.\\
\end{multline}
As this parameterizes the real unit circle over a support of two orthonormal states, there is some value of $\theta_1$ that gives any real combination of the two vectors. If we take the case $W=0\neq V$ then we can get an expression for the angle to get eigenstates $\ket{1,0,0;j}$ using Equation~(\ref{genang}) below.

$N=2, M=1$ is a special case because the eigenstate coefficients are independent of $W$. The coefficients are also real for any real $V$. Therefore the $N=2$ eigenstates are real in the trigonometric, rational and hyperbolic cases.

For $M=1$ the relationship between $\Theta_1$ and the spectral parameter $E_1$ may be obtained without assuming that $N$ is even. Equating Equations~(\ref{e1eq}) and~(\ref{out1}) gives the relationship between $\theta_1$ and $E_1$:
\begin{equation}
    \sin\frac{\theta_1}{2}=\frac{E_1+1}{\sqrt{2\left(E_1^2+1\right)}}.
\end{equation}
We can obtain the relationship between $\theta_1$ and $V$ in the case $W=0$:
\begin{equation}\label{genang} 
    \cos\frac{\theta_1}{2}=\left[1+\frac{\left(N-\sqrt{N^2+3^{N-2}V^2}\right)^2}{3^{N-2}V^2}\right]^{-1/2}.
\end{equation}

\subsection{$\mathbf{N=3}$ Eigenstates}\label{demoofn3}
For $N=3$ particles, $M=0$ is forbidden by Equation~(\ref{ndef}) and for $M=1$ we have fiducial states $\ket{0,1}$ and $\ket{1,0}$. We have two Bethe Equations:
\begin{equation}\label{n3eqs1}
    1-\frac{\eta}{3\left(E_1^2-\eta^2\right)}\left(-3gs E_1^2+4VE_1-3g\right)=0
\end{equation}
for fiducial state $\ket{0,1}$ and
\begin{equation}\label{n3eqs2}
    1-\frac{\eta}{3\left(E_1^2-\eta^2\right)}\left(3gsE_1^2+4VE_1+3g\right)=0
\end{equation}
for fiducial state $\ket{1,0}$. Note that Equations~(\ref{n3eqs1}) and~(\ref{n3eqs2}) are quadratic. The resulting eigenstates are
\begin{subequations}\label{n3states}
    \begin{equation}
        \ket{1,0,1;0}=\gamma_0\left[\frac{-3+W-F_-}{\sqrt{3}V}\ket{2,1}+\ket{0,3}\right],
    \end{equation}
    \begin{equation}
        \ket{1,0,1;1}=\gamma_1\left[\frac{-3+W+F_-}{\sqrt{3}V}\ket{2,1}+\ket{0,3}\right],
    \end{equation}
    \begin{equation}
        \ket{1,1,0;2}=\gamma_2\left[-\frac{3+W+F_+}{\sqrt{3}V}\ket{3,0}+\ket{1,2}\right],
    \end{equation}
    \begin{equation}
        \ket{1,1,0;3}=\gamma_3\left[\frac{-3-W+F_+}{\sqrt{3}V}\ket{3,0}+\ket{1,2}\right],
    \end{equation}
\end{subequations}
where
\begin{equation}\label{fdef}
        F_\pm\equiv \sqrt{9+3V^2\pm 6W+W^2}.
\end{equation}

The normalization factors $\gamma_i$ are given in Appendix~\ref{n3app}. As the $M=1$ EGO circuit outputs cover the unit circle over the support, any of these real-coefficient LMG eigenstates can be generated by the circuit given in Equation~(\ref{logego}). As for the $N=2,M=1$ case there is no relationship between $W$ and $V$ that could lead to complex eigenstates. Therefore any trigonometric or hyperbolic $N=3$ eigenstate can be created by the circuit~(\ref{logego}).

\subsection{$\mathbf{N=4}$ Eigenstates}
For $N=4$ eigenstates Equation~(\ref{ndef}) allows $M=1$ while $\nu_a=\nu_b=1$ and $M=2$ while $\nu_a=\nu_b=0$. The $M=1$ eigenstates may be obtained in the same way as $N=3$ in Subsection~\ref{demoofn3}. To generate the unnormalized $M=2$, any-$N$ eigenstates we use Equation~(\ref{ego}):
\begin{multline}\label{m2un} 
    \ket{\text{EGO}_{2,\nu_a,\nu_b}}=\frac{\sqrt{6\left(3-2\left(-1\right)^{\nu_a}\right)}}{\left(E_1+\eta\right)\left(E_2+\eta\right)}\ket{\nu_a+4,\nu_b}\\
    +\frac{\sqrt{6\left(3-2\left(-1\right)^{\nu_b}\right)}}{\left(E_1-\eta\right)\left(E_2-\eta\right)}\ket{\nu_a,\nu_b+4}\\
    +3^\frac{\nu_a+\nu_b}{2}\left(\sum_\pm\left(E_1\pm \eta\right)^{-1}\left(E_2\mp \eta\right)^{-1}\right)\ket{\nu_a+2,\nu_b+2}.\\
\end{multline}
The Bethe equations for $M=2,N=4$ and therefore fiducial state $\ket{0,0}$ from Equation~(\ref{beqs}) are

\begin{subequations}\label{n4eqs}
\begin{equation}
    1-\frac{V E_1\eta}{2\left(E_1^2-\eta^2\right)}+2g\frac{1+s E_1E_2}{E_1-E_2}=0,
\end{equation}
\begin{equation}
    1-\frac{V E_2\eta}{2\left(E_2^2-\eta^2\right)}+2g\frac{1+s E_2E_1}{E_2-E_1}=0.
\end{equation}
\end{subequations}

Equation~(\ref{m2ell}) yields a solution for the simplified $W=0\neq V$ case:
\begin{equation}\label{m2specs}
    E_1=\frac{V-\sqrt{V^2+16}}{4}\quad,\quad E_2=\frac{V+\sqrt{V^2+16}}{4}.
\end{equation}
To generate all of the $M=2$ eigenstates, apply the solutions of Equations~(\ref{n4eqs}) to the EGO~(\ref{ego}) and normalize as before with $N=2,3$.

The logarithmic-depth EGO circuit~(\ref{logego}) outputs
\begin{multline}
    \ket{\psi_2}=\cos\frac{\theta_1}{2}\sin\frac{\theta_2}{2}\ket{2^0}_3\\
    +\sin\frac{\theta_1}{2}\ket{2^1}_3+\cos\frac{\theta_1}{2}\cos\frac{\theta_2}{2}\ket{2^2}_3\\
\end{multline}
which parameterizes the $2$-sphere over the specified support. Thus any $M=2$ trigonometric LMG eigenstate can be realized for some set of values $\Theta_2$, as any real unit $3$-vector can be mapped to a point on the $2$-sphere.

We may relate $E_1,E_2$ to the logarithmic-depth circuit~(\ref{logego}) rotation angles $\Theta_2$ in the simplified case of $W=0$ and $N$ even by using Equation~(\ref{m2ell}). Upon normalization this yields an eigenstate which can be compared to the $M=2$ output of the logarithmic circuit; one component of said circuit has a single $\sin\theta_1/2$ as a coefficient. Depending on the bosonic-to-computational-state encoding this can be mapped to a single component of the eigenstate in terms of the EGO~(\ref{ego}), which can give an expression for $\theta_1$. The second angle can be determined using either of the other two components.

\subsection{Eigenstate Preparation for $\mathbf{M>2}$}
For quintic or higher polynomials no general analytic solution exists for all cases~\cite{abel_2012}, so the Bethe Equations~(\ref{beqs}) must be solved numerically beyond $M=2$.

Each output vector of either the logarithmic- or linear-depth EGO circuit~(\ref{logego}) $\ket{\psi_M}$ is a state with $(M+1)$ nonzero real amplitudes. Each amplitude is a product of at most $M$ sine or cosine functions of a single half-angle. Given an eigenstate, then, the EGO circuit's~(\ref{logego}) angles of rotation can be obtained numerically or using Equation~(\ref{linearangles}) in the linear case. Any real unit vector over the basis $\{\ket{2^0}_{M+1},\ket{2^1}_{M+1},...,\ket{2^M}_{M+1}\}$ can be prepared with either circuit. 


Appendix~(\ref{excircs}) gives two explicit examples for preparing the ground state of the $M=3,\text{ }N=7$-particle and $M=10,\text{ } N=20$-particle LMG models for both the logarithmic- and linear-depth circuits~(\ref{logego}).

\section{Discussion}\label{discsec}

By utilizing the product nature of the EGO~(\ref{ego}) we have constructed two $\mathcal{O}(N)$-gate quantum circuit analogues that function as variational ansatzes for the trigonometric LMG model of $N$ particles. Both circuits are described by Equation~(\ref{logego}) while the logarithmic-depth version is also shown in Figure~(\ref{linfig}). In the case of the linear-depth version, the input angles are described by Equation~(\ref{linearangles}). Low qubit count, depth, gate count and parallelization makes the circuits well-suited for NISQ devices. The circuits will work for hyperbolic or rational cases of the LMG such that $E_\ell\in\mathbb{R}\text{ }\forall\text{ }1\leq \ell\leq M$, of which there are at least three cases: $N=1,2,3$. No single phase transition distinguishes trigonometric cases with purely real spectral parameters from those with complex ones \cite{LermaH.:2013cla}. For $M\geq 3$ the Bethe equations~(\ref{beqs}) are quintic and lack analytic solutions but $E_\ell$ can be solved numerically.

As an exactly-solvable model, the LMG model can be used to evaluate near-term quantum devices. By comparing results obtained from exact solutions to NISQ-device simulations the LMG model can function as a benchmark, just as it did for classical calculations \cite{LIPKIN1965188}.

These circuits will function as a variational ansatz for any linear-support, real-eigenstate model. The greater family of bosonic Gaudin models would be fertile ground for further investigations~\cite{LermaH.:2013cla}. The circuits we describe are also able to create arbitrary real superpositions of LMG eigenstates for a given $M,\nu_a,\nu_b$ as the circuit's output vector parameterizes the unit $M$-sphere.

\section{Acknowledgements}
This work was supported by DOE award number DE-SC0019465. K.R. thanks Will Kirby, Michael Kreschuk, Andrew Tranter, Alan Ottenstein and users on StackExchange for useful discussions.

\appendix 
\section{$\mathbf{N=3}$ Normalization Factors}\label{n3app}

The normalization factors from Equations~(\ref{n3states}) are
\begin{subequations}
\begin{equation}
    \gamma_0=\left(1+\left(\frac{3-W+F_-}{\sqrt{3}V}\right)^2\right)^{-1/2},
\end{equation}
\begin{equation}
    \gamma_1=\left(1+\left(\frac{-3+W+F_-}{\sqrt{3}V}\right)^2\right)^{-1/2},
\end{equation}
\begin{equation}
    \gamma_2=\left(1+\left(\frac{3+W+F_+}{\sqrt{3}V}\right)^2\right)^{-1/2},
\end{equation}
\begin{equation}
    \gamma_3=\left(1+\left(\frac{-3-W+F_+}{\sqrt{3}V}\right)^2\right)^{-1/2}
\end{equation}
\end{subequations}
where $F_\pm$ is defined in Equation~(\ref{fdef}).

\section{Qudit Circuit}
\begin{figure*} 
\begin{center}
\includegraphics[scale=0.43]{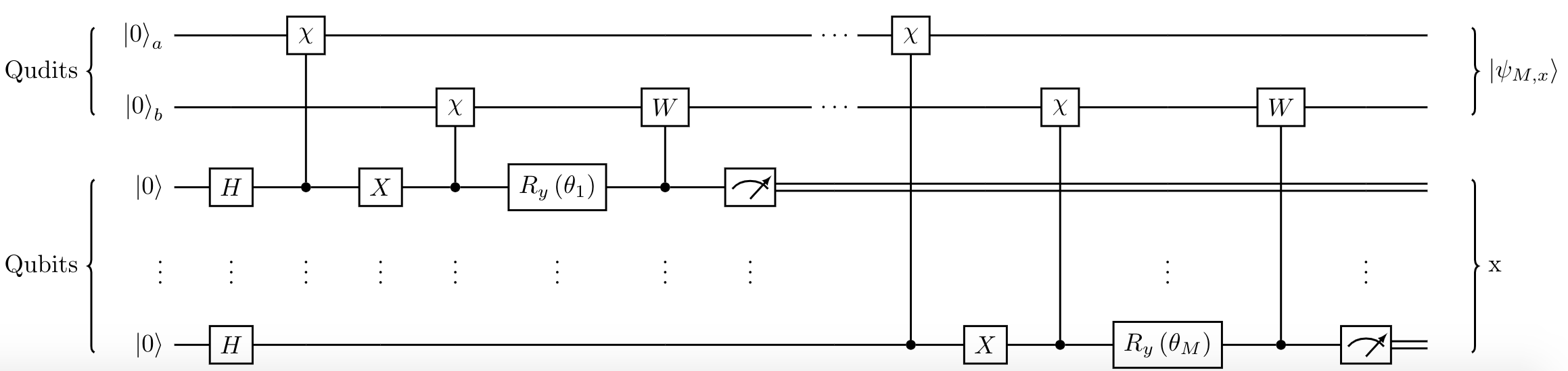} 
\caption{A $2$-qudit ($d=M+1=\left(N-\nu_a-\nu_b+2\right)/2$), $M$-ancilla circuit for the Lipkin-Meshkov-Glick (LMG) Eigenstate Generating Operator (EGO) factor from Equation~(\ref{ego}). The two qudits represent bosonic modes $a$ and $b$. The number of ancilla $M$ scales as $N/2$. Gates $\chi$ and $W$ are given in Equation~(\ref{wchi}). Each application of the circuit requires $M$ measurements which generate a measurement bitstring $x$ associated with the resultant qudit state $\ket{\psi_{M,x}}$. We demonstrate that for any $M$ (and by extension any number of particles $N$) any eigenstate of the trigonometric LMG is equal to at least one output state $\ket{\psi_{M,x}}$ for some $x$ and some set of values for the rotation parameters $\Theta_M$. In the context of benchmarking the Variational Quantum Eigensolver (VQE), a valid ansatz for any LMG ground state can be prepared using this circuit. VQE could then be applied and its results compared to exactly-known solutions from the LMG's underlying $XXZ$ Gaudin algebraic structure.}
\label{egofig}
\end{center}
\end{figure*}

Here we present a probabilistic qudit-based EGO circuit shown in Figure~\ref{egofig}. It requires two qudits with dimension $d=M+1$ as well as $M$ ancilla qubits. The qudits encode bosonic modes with a maximum occupancy of $M$. By entangling the qudits with the ancilla qubits, acting on the ancilla qubits with rotation gates, and measuring the ancillae we obtain an output state $\ket{\psi_{M,x}}$ and a labeling bitstring $x$ of length $M$.

There are two single-qudit gates which appear in the circuit:
\begin{equation}\label{wchi}
    \chi\ket{k}\equiv\ket{k+1\text{ mod }M+1}\quad W\ket{k}\equiv(-1)^{1-\delta_{k0}}\ket{k}
\end{equation}
where $k$ is an integer from $0$ to $M$. The circuit also uses the Hadamard, $R_y$ and Pauli-$X$ gates on the ancilla qubits.

Because $M$ measurements are performed, there are $2^M$ possible output states. It is not guaranteed that the target eigenstate is preparable with every output. If $p_M$ is the probability of the circuit successfully preparing the target eigenstate, this qudit-based circuit has a gate count $6M/p_M$ as opposed to the linear- and logarithmic-circuits' $M$. In having two qudits functioning as effective bosonic modes and $\chi$ acting as effective creation-squared operators this circuit is more directly based on the EGO~(\ref{ego}). It is included to demonstrate the use of qudits to simulate bosonic systems.

There is a recursion relation between outputs of the qudit-based circuit with different iterative depths. If the output of the qudit-based circuit iterated $t$ times where  bitstring $x$ was measured is
\begin{equation}
    \ket{\psi_{t}}=\sum_{k=0}^tc_k\ket{2t+\nu_a-2k,\nu_b+2k}
\end{equation}
then the output state after iterating the circuit one more time is
\begin{multline}
    \ket{\psi_{t+1}}=\sum_{j=0}^t\sum_{k=0}^{t-1}\frac{c_j}{\sqrt{1-\sin\theta_{t+1}c_kc_{k+1}}}\times\\
    \Big(\cos\frac{\theta_{t+1}}{2}\ket{2t-2j+\nu_a+2,2j+\nu_b}\\
    -\sin\frac{\theta_{t+1}}{2}\ket{2t-2j+\nu_a,2j+2+\nu_b}\Big)
\end{multline}
if the $(t+1)\textsuperscript{th}$ measurement yields $0$ and
\begin{multline}
    \ket{\psi_{t+1}}=\sum_{j=0}^t\sum_{k=0}^{t-1}\frac{c_j}{\sqrt{1+\sin\theta_{t+1}c_kc_{k+1}}}\times\\
    \Big(\left(-1\right)^{1-\delta_{j0}}\sin\frac{\theta_{t+1}}{2}\ket{2t-2j+2+\nu_a,2j+\nu_b}\\
    -\cos\frac{\theta_{t+1}}{2}\ket{2t-2j+\nu_a,2j+2+\nu_b}\Big)\\
\end{multline}
if the $(t+1)\textsuperscript{th}$ measurement yields $1$.

\section{Logarithmic and Linear Circuit Output States}\label{logputs}
We include some of the output states for both circuits described in the main body of this paper. We use the previous notation that $\ket{n}_m$ will represent a bitstring of length $m$ and representing binary integer $0\leq n\leq 2^m-1$. State $\ket{\psi_M}$ is the output of the circuit given in Figure~\ref{linfig} and Equation~(\ref{logego}) after $M$ iterations. Note that every nonzero vector component is a product of at most $M$ sine or cosine functions which are functions of exactly one half-angle. Further, no angle appears twice in a component. The logarithmic-depth circuit has outputs
\begin{subequations}
\begin{equation}
    \ket{\psi_0}=\ket{2^0}_1\\
\end{equation}
\begin{equation}
    \ket{\psi_1}=\sin\frac{\theta_1}{2}\ket{2^0}_2+\cos\frac{\theta_1}{2}\ket{2^1}_2 \\
\end{equation}
\begin{multline}
    \ket{\psi_2}=\cos\frac{\theta_1}{2}\sin\frac{\theta_2}{2}\ket{2^0}_3\\
    +\sin\frac{\theta_1}{2}\ket{2^1}_3+\cos\frac{\theta_1}{2}\cos\frac{\theta_2}{2}\ket{2^2}_3\\
\end{multline}
\begin{multline}
    \ket{\psi_3}=\sin\frac{\theta_1}{2}\sin\frac{\theta_3}{2}\ket{2^0}_4+\cos\frac{\theta_1}{2}\sin\frac{\theta_2}{2}\ket{2^1}_4\\
    +\sin\frac{\theta_1}{2}\cos\frac{\theta_3}{2}\ket{2^2}_4+\cos\frac{\theta_1}{2}\cos\frac{\theta_2}{2}\ket{2^3}_4\\
\end{multline}
\begin{multline}
    \ket{\psi_4}=\cos\frac{\theta_1}{2}\cos\frac{\theta_2}{2}\sin\frac{\theta_4}{2}\ket{2^0}_5+\sin\frac{\theta_1}{2}\sin\frac{\theta_3}{2}\ket{2^1}_5\\
    +\cos\frac{\theta_1}{2}\sin\frac{\theta_2}{2}\ket{2^2}_5+\sin\frac{\theta_1}{2}\cos\frac{\theta_3}{2}\ket{2^3}_5\\
    +\cos\frac{\theta_1}{2}\cos\frac{\theta_2}{2}\cos\frac{\theta_4}{2}\ket{2^4}_5\\
\end{multline}
\begin{multline}
    \ket{\psi_5}=\sin\frac{\theta_1}{2}\cos\frac{\theta_3}{2}\sin\frac{\theta_5}{2}\ket{2^0}_6\\
    +\cos\frac{\theta_1}{2}\cos\frac{\theta_2}{2}\sin\frac{\theta_4}{2}\ket{2^1}_6+\sin\frac{\theta_1}{2}\sin\frac{\theta_3}{2}\ket{2^2}_6\\
    +\cos\frac{\theta_1}{2}\sin\frac{\theta_2}{2}\ket{2^3}_6+\sin\frac{\theta_1}{2}\cos\frac{\theta_3}{2}\cos\frac{\theta_5}{2}\ket{2^4}_6\\
    +\cos\frac{\theta_1}{2}\cos\frac{\theta_2}{2}\cos\frac{\theta_4}{2}\ket{2^5}_6\\
\end{multline}
\end{subequations}
and the linear-depth circuit has outputs
\begin{subequations}
\begin{equation}
    \ket{\psi_0}=\ket{2^0}_1
\end{equation}
\begin{equation}
    \ket{\psi_1}=\sin\frac{\theta_1}{2}\ket{2^0}_2+\cos\frac{\theta_1}{2}\ket{2^1}_2
\end{equation}
\begin{multline}
    \ket{\psi_2}=\sin\frac{\theta_1}{2}\sin\frac{\theta_2}{2}\ket{2^0}_3+\sin\frac{\theta_1}{2}\cos\frac{\theta_2}{2}\ket{2^1}_3\\
    +\cos\frac{\theta_1}{2}\ket{2^2}_3
\end{multline}
\begin{multline}\label{m3lin}
    \ket{\psi_3}=\sin\frac{\theta_1}{2}\sin\frac{\theta_2}{2}\sin\frac{\theta_3}{2}\ket{2^0}_4\\
    +\sin\frac{\theta_1}{2}\sin\frac{\theta_2}{2}\cos\frac{\theta_3}{2}\ket{2^1}_4\\
    +\sin\frac{\theta_1}{2}\cos\frac{\theta_2}{2}\ket{2^2}_4+\cos\frac{\theta_1}{2}\ket{2^3}_4
\end{multline}
\begin{multline}
    \ket{\psi_4}=\sin\frac{\theta_1}{2}\sin\frac{\theta_2}{2}\sin\frac{\theta_3}{2}\sin\frac{\theta_4}{2}\ket{2^0}_5\\
    +\sin\frac{\theta_1}{2}\sin\frac{\theta_2}{2}\sin\frac{\theta_3}{2}\cos\frac{\theta_4}{2}\ket{2^1}_5\\
    +\sin\frac{\theta_1}{2}\sin\frac{\theta_2}{2}\cos\frac{\theta_3}{2}\ket{2^2}_5\\
    +\sin\frac{\theta_1}{2}\cos\frac{\theta_2}{2}\ket{2^3}_5+\cos\frac{\theta_1}{2}\ket{2^4}_5
\end{multline}
\begin{multline}
    \ket{\psi_5}=\sin\frac{\theta_1}{2}\sin\frac{\theta_2}{2}\sin\frac{\theta_3}{2}\sin\frac{\theta_4}{2}\sin\frac{\theta_5}{2}\ket{2^0}_6\\
    +\sin\frac{\theta_1}{2}\sin\frac{\theta_2}{2}\sin\frac{\theta_3}{2}\sin\frac{\theta_4}{2}\cos\frac{\theta_5}{2}\ket{2^1}_6\\
    +\sin\frac{\theta_1}{2}\sin\frac{\theta_2}{2}\sin\frac{\theta_3}{2}\cos\frac{\theta_4}{2}\ket{2^2}_6\\
    +\sin\frac{\theta_1}{2}\sin\frac{\theta_2}{2}\cos\frac{\theta_3}{2}\ket{2^3}_6\\
    +\sin\frac{\theta_1}{2}\cos\frac{\theta_2}{2}\ket{2^4}_6+\cos\frac{\theta_1}{2}\ket{2^5}_6.
\end{multline}
\end{subequations}
\section{}\label{logproofsec}
The linear-depth EGO circuit given in Equation~(\ref{logego}) produces output states with nonzero amplitudes parameterized by $(M+1)$ dimensional hyperspherical coordinates. Thus states generated by these cicuits are in correspondence with points on the $M$-sphere~\cite{nspherestuff}. Hence these circuits can generate any quantum state with real amplitudes and given support.  The logarithmic-depth circuit generates states whose nonzero amplitudes are not parameterized by hyperspherical coordinates. In this section we prove that nonetheless these circuits generate states in one-to-one correspondence with points on the $M$ sphere.

We proceed by induction to prove that the output vector of the logarithmic-depth circuit given by Equation~(\ref{logego}) and Figure~\ref{linfig} can parameterize the $M$-sphere over the set of bitstring vectors with support $M+1$ and Hamming weight $1$.

The output of Equation~(\ref{logego}) for $M=1$, given by Equation~(\ref{out1}), parameterizes a real unit circle ($1$-sphere) over the support $\ket{2^0}_2$ and $\ket{2^1}_2$. Thus the output will parameterize any real linear combination of computational basis vectors with length $2$ and Hamming weight $1$. Projecting the parameterization $\ket{\psi_1}$ onto the hyperplane $H_1$ defined by $\ket{2^1}_2$ will result in a parameterization of the $1$-ball (line segment from $-1$ to $1$) on $H_1$. This will function as our base case for a proof by induction.

As an inductive hypothesis, assume that output vector $\ket{\psi_t}$ parameterizes the $t$-sphere over the support $\{\ket{2^0}_{t+1},\ket{2^1}_{t+1},...,\ket{2^{t}}_{t+1}\}$. Define $\ket{\psi_t'}$ as $\ket{\psi_t}$ projected onto the hyperplane $H_t$ defined by $\{\ket{2^1}_{t+1},...,\ket{2^{t}}_{t+1}\}$ and assume that $\ket{\psi_t'}$ parameterizes the $t$-ball.

By definition, a $t$-ball is the hypervolume contained within a $(t-1)$-sphere. We may also restate this as the hypervolume contained in between a $t$-ball and a $t$-sphere; in the base case these were the line segment $-1$ to $1$ and the $1$-sphere (i.e. circle). The $t$-ball is parameterized if $\ket{\psi_t'}$ can parameterize any real linear combination of basis vectors given by $H_t$ where the sum of the squared coefficients is less than or equal to $1$. 

We can write output vectors $\ket{\psi_t}$ and $\ket{\psi_{t+1}}$ as 
\begin{align}\label{psites}
    &\ket{\psi_t}=\sum_{j=0}^tc_{j+1}\ket{2^j}_{t+1}\\
    &\ket{\psi_{t+1}}=\sum_{j=0}^{t+1}d_{j+1}\ket{2^j}_{t+2}
\end{align}
where $c_{j+1},d_{j+1}$ are real, normalized coefficients. Note that this requires $d_1=\pm\sqrt{1-\sum_{j=1}^{t+1}d_{j+1}^2}$.

To act on this state with another factor of the EGO circuit, a $(t+2)$\textsuperscript{th} qubit in state $\ket{0}$ must be appended onto the system as a target for a controlled-$R_y\left(\theta_{t+1}\right)$ gate with its control on qubit $1\leq m <t+2$.

The support of $\ket{\psi_t}\otimes\ket{0}$ has Hamming weight $1$. Thus when we act on $\ket{\psi_t}\otimes\ket{0}$ with $C^m\left(R_y\left(\theta_{t+1}\right)_{t+2}\right)$ the result is
\begin{multline}
   C^m\left(R_y\left(\theta_{t+1}\right)_{t+2}\right)\ket{\psi_t}\otimes\ket{0} = \sum_{j=0}^tc_{j+1}\ket{2^j}_{t+1}\otimes\\
    \Bigg(\ket{0}+\Bigg(\left(\cos\frac{\theta_{t+1}}{2}-1\right)\ket{0}+\sin\frac{\theta_{t+1}}{2}\ket{1}\Bigg)\delta_{j+1,m}\Bigg).\\
\end{multline}

This CNOT gate has its target on qubit $m$ and control on qubit $(t+2)$. The only vector in the support with $\ket{1}$ as the final qubit also has $\ket{1}$ as qubit $m$. Therefore $\ket{\psi_{t+1}}$ is given by:
\begin{multline}\label{cyp1}
    \ket{\psi_{t+1}}=c_m\left(\cos\frac{\theta_{t+1}}{2}\ket{2^m}_{t+2}+\sin\frac{\theta_{t+1}}{2}\ket{2^0}_{t+2}\right)\\
    +\sum_{j=0}^tc_{j+1}\left(1-\delta_{j+1,m}\right)\ket{2^{j+1}}_{t+2}.\\
\end{multline}

To demonstrate that $\ket{\psi_{t+1}}$ covers the $(t+1)$-sphere over the proper support, examine $\ket{\psi_{t+1}}$ with $\theta_{t+1}$ set to $0$. This returns $\ket{\psi_{t+1}}$ to the form of $\ket{\psi_{t}}$
\begin{equation}
    \ket{\psi_{t+1}}\Big|_{\theta_{t+1}\rightarrow 0}=\sum_{j=0}^tc_{j+1}\ket{2^{j+1}}_{t+2}
\end{equation}
which covers the $t$-sphere on hyperplane $H_{t+1}$ by assumption. If we set $\theta_{t+1}\rightarrow \pi$ we obtain
\begin{multline}\label{psiproj2}
    \ket{\psi_{t+1}}\Big|_{\theta_{t+1}\rightarrow\pi}=c_m\ket{2^0}_{t+2}\\
    +\sum_{j=0}^tc_{j+1}(1-\delta_{j+1,m})\ket{2^{j+1}}_{t+2}.
\end{multline}

When projected onto $H_{t+1}$, Equation~(\ref{psiproj2}) parameterizes the $t$-ball by assumption. The projected vector $\ket{\psi_{t+1}'}$ is continuous with respect to $\theta_{t+1}$ so $\ket{\psi_{t+1}'}$ parameterizes the space between the two surfaces at $\theta_{t+1}\rightarrow 0$ and $\theta_{t+1}\rightarrow\pi$. Thus $\ket{\psi_{t+1}'}$ parameterizes the $(t+1)$-ball on $H_{t+1}$.

This means that $\ket{\psi_{t+1}'}$ parameterizes any combination of $\{d_2,...,d_{t+2}\}$ such that $\sum_{j=2}^{t+2}d_j^2\leq 1$. We also know that $\ket{\psi_{t+1}}$ is a unit vector. Thus $\ket{\psi_{t+1}}$ parameterizes any combination of $\{d_1,...,d_{t+1}\}$ up to a possible factor of $-1$ on $d_1$. Therefore if output vector $\ket{\psi_{t}}$ parameterizes the $t$-sphere and its projection parameterizes the $t$-ball, output vector $\ket{\psi_{t+1}}$ parameterizes the $(t+1)$-sphere up to a possible factor of $-1$ on the $\ket{2^0}_{t+2}$ component.

Equation~(\ref{psites}) indicates that $\ket{d_1}$ is the scalar multiple of basis vector $\ket{2^0}_{t+2}$ for $\ket{\psi_{t+1}}$. Equation~(\ref{cyp1}) shows that $\ket{2^0}_{t+2}$ is multiplied by $\sin\theta_{t+1}/2$ while no other basis vectors are. Therefore letting $\theta_{t+1}\rightarrow -\theta_{t+1}$ changes the sign of only the $\ket{2^0}_{t+2}$ component of $\ket{\psi_{t+1}}$. In the context of the hyperplane, letting $\theta_{t+1}$ go to $-\theta_{t+1}$ constitutes an inversion over $H_{t+1}$.

We've already established that $\ket{\psi_{t+1}}$ parameterizes the $(t+1)$-sphere up to a possible factor of $-1$ on the $\ket{2^0}_{t+2}$ component; thus simultaneously being able to uniquely negate that factor means that the $(t+1)$-sphere is completely parameterized by $\ket{\psi_{t+1}}$. Therefore if output vector $\ket{\psi_{t}}$ parameterizes the $t$-sphere, output vector $\ket{\psi_{t+1}}$ parameterizes the $(t+1)$-sphere.

\section{Example Circuits}\label{excircs}
In this section we provide examples of using an $M=3$ and $M=10$ circuit defined by Equation~(\ref{logego}) to prepare the ground state of the $N=7$ and $N=20$ LMG ground states. We take $W=1/2$ and $V=3/4$ in the trigonometric LMG regime, which forces $\eta=-\sqrt{5}$. ``Linear-depth circuit" refers to Equation~(\ref{logego}) where $f(n)$ is defined by the upper case of Equation~(\ref{fnseq}) and ``logarithmic-depth circuit" refers to Equation~(\ref{logego}) when $f(n)$ is defined by the lower case.

\subsection{$\mathbf{N=7}$ Example Circuit}

We begin by solving the quintic $M=3$ Bethe equations~(\ref{beqs}) numerically. Once we solve the equations for $M=3,\nu_a=0,\nu_b=1$ and $M=3,\nu_a=1,\nu_b=0$ we input the spectral parameter solutions into the eigenvalue Equation~(\ref{evaleq}) to determine which parameters give the ground state. For $N=7$, $W=1/2$, $V=3/4$ the spectral parameters which yield the lowest expectation value are
\begin{align*}
    E_1 &= 1.94591\\
    E_2 &= 1.33363\\
    E_3 &=0.701066
\end{align*}
while $\nu_a=1$ and $\nu_b=0$. Now that the variables in the EGO~(\ref{ego}) have been defined we may directly calculate and normalize the ground state:
\begin{multline}
    \ket{3,1,0;1}=3.40577\times10^{-3}\ket{1,6}-3.08911\times 10^{-2}\ket{3,4}\\
    +0.18121 \ket{5,2}-0.982953\ket{7,0}\\
\end{multline}
which we encode via Equation~(\ref{sampleencoding})
\begin{multline}
    \ket{3,1,0;1}= 3.40577\times10^{-3}\ket{1000}\\
   -3.08911\times 10^{-2}\ket{0100}+0.18121 \ket{0010}-0.982953\ket{0001}.\\
\end{multline}

\begin{figure*} 
\begin{center}
\includegraphics[scale=0.46]{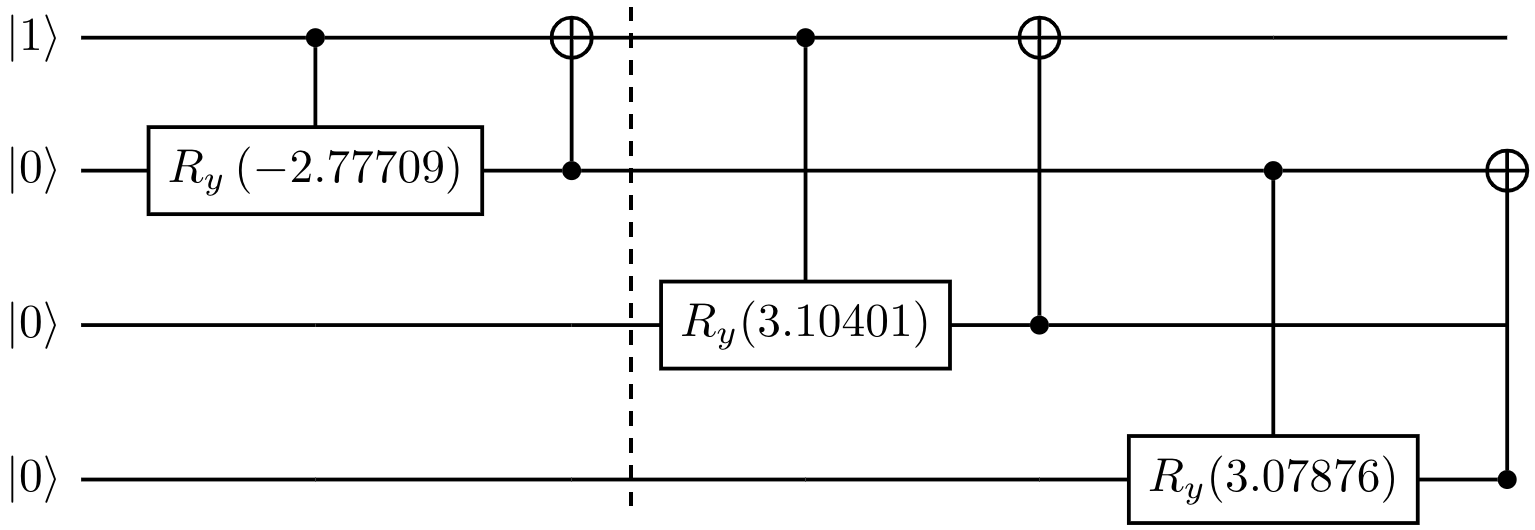} 
\caption{A $4$-qubit, $6$-gate circuit which prepares the ground state of the $N=7$-particle Lipkin-Meshkov-Glick (LMG) Hamiltonian when interaction parameters $V,W$ are equal to $3/4$ and $1/2$ respectively. The circuit has depth $2$ because the second and third pairs of gates can be run simultaneously. A generalized form of the circuit for any $N$ is given by Equation~(\ref{logego}).}
\label{excirc7}
\end{center}
\end{figure*}

Equation~(\ref{linearangles}) gives the linear-depth circuit's angles as
\begin{align*}
    \theta_1 &= 3.13478\\
    \theta_2 &= 3.20338\\
    \theta_3 &=9.78939
\end{align*}
To check the accuracy of such a process, we truncate the above angles to $6$ decimal places, substitute them into the $M=3$ linear-depth circuit output in Equation~(\ref{m3lin}), and compute the expectation value of the LMG Hamiltonian~(\ref{bham}) in units of the energy gap $\epsilon$. This can be compared to the ground state energy predicted by the Bethe ansatz with Equation~(\ref{evaleq}): $-3.34051529185\epsilon$ for the calculated state's expectation value and $-3.34051529181\epsilon$ for the ground state energy from Equation~(\ref{evaleq}). The relative error between the two values is $1.2\times10^{-11}\epsilon$.

We may also find the logarithmic-depth circuit's angles numerically
\begin{align*}
    \theta_1 &= -2.77709\\
    \theta_2 &= 3.10401\\
    \theta_3 &= 3.07876
\end{align*}
which yield a ground state approximation with an expectation value of $-3.340515291813\epsilon$. This corresponds to a relative error of $4.4\times 10^{-12}\epsilon$. The logarithmic circuit to prepare the ground state of the $N=7$ LMG model with $V=3/4$ and $W=1/2$ is shown in Figure~\ref{excirc7}.


\subsection{$\mathbf{N=20}$ Example Circuit}

To illustrate an alternative way to construct the LMG eigenstates, one can  take the Hamiltonian in angular momentum representation in Equation~(\ref{angham}) and numerically diagonalize the $j=10$ block, then translate back into the bosonic representation. The resulting ground state is then given by
\begin{multline}
    \ket{10,0,0;1}=5.73265\times 10^{-10}\ket{0,20} -2.22883\times 10^{-8}\ket{2,18}\\
    +3.72394\times 10^{-7}\ket{4,16} -4.12631\times 10^{-6}\ket{6,14}\\
    +3.49942\times 10^{-5}\ket{8,12}-2.4413\times 10^{-4}\ket{10,10}\\
    +1.47154\times 10^{-3}\ket{12,8}-7.89635\times 10^{-3}\ket{14,6}\\
    +0.0389319\ket{16,4}-0.184149\ket{18,2}+0.982094\ket{20,0}.
\end{multline}

We may then substitute this into our encoding Equations~(\ref{sampleencoding}) and~(\ref{linearangles}) to get the angles of rotation for the linear-depth circuit
\begin{align}\label{20angs}
    &\theta_1 = 3.14159 \quad &\theta_2 = 3.14159\\
    &\theta_3 =3.14159 \quad &\theta_4=3.14160\nonumber \\
    &\theta_5=3.14152 \quad &\theta_6=3.14208\nonumber\\
    &\theta_7=3.13865 \quad &\theta_8=3.15739 \nonumber\\
    &\theta_9=3.06371 \quad &\theta_{10}=3.51230\nonumber.
\end{align}

The first $4$ angles of Equation~(\ref{20angs}) are noticeably close to $\pi$ but they are not equal to $\pi$ according to Equation~(\ref{linearangles}). For $j=1,2,3,4$ the relevant coefficients from $\ket{10,0,0;1}$ are
\begin{align}
    &c_8=-4.12631\times10^{-6}\\
    &c_9=3.72394\times10^{-7}\nonumber\\
    &c_{10}=-2.22883\times10^{-8}\nonumber\\
    &c_{11}=5.73265\times10^{-10}\nonumber.
\end{align}

Because these coefficients are all much smaller than unity, we know that $\sum_{k=1}^{M+2-j}c_k^2\approx 1$ for low $j$. Thus we may linearize Equation~(\ref{linearangles}) to obtain
\begin{equation}\label{ultralin}
    \theta_j\approx \pi-2c_{M+2-j}
\end{equation}
for low $j$ in this case. The linearized Equation~(\ref{ultralin}) agrees with the angles given in Equation~(\ref{20angs}) for low $j$.

\begin{figure*} 
\begin{center}
\includegraphics[width=9cm,height=19.1cm]{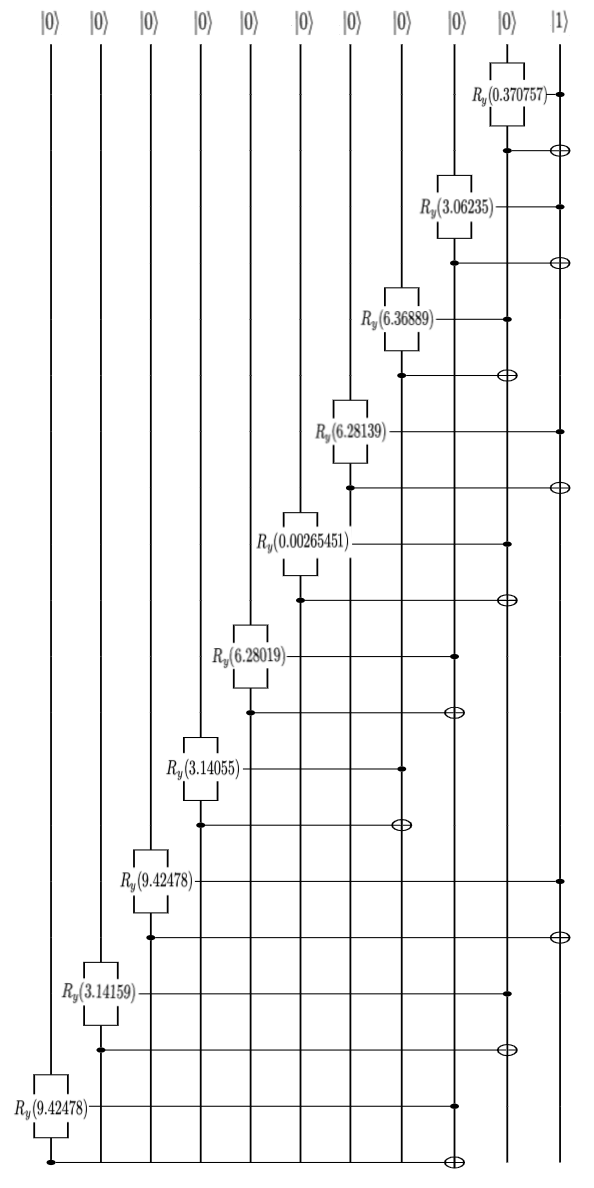} 
\caption{An $11$-qubit, $20$-gate circuit which prepares the ground state of the $N=20$-particle Lipkin-Meshkov-Glick (LMG) Hamiltonian when interaction parameters $V,W$ are equal to $3/4$ and $1/2$ respectively. The general form of the circuit to prepare an $N$-particle LMG eigenstate is given in Equation~(\ref{logego}). The dotted lines are to distinguish sections were gates can be run in parallel.}
\label{excirc20}
\end{center}
\end{figure*}

The logarithmic-depth circuit's angles, shown in Figure~\ref{excirc20}, can be obtained numerically:
\begin{align}
    &\theta_1 = 0.370757 \quad &\theta_2 = 3.06235\\
    &\theta_3 =6.36889 \quad &\theta_4=6.28139\nonumber \\
    &\theta_5=2.65451\times 10^{-3} \quad &\theta_6=6.28019\nonumber\\
    &\theta_7=3.14055 \quad &\theta_8=9.42478 \nonumber\\
    &\theta_9=3.14159 \quad &\theta_{10}=9.42478\nonumber.
\end{align}

\bibliographystyle{unsrtnat} 
\bibliography{references}
\end{document}